\documentclass[aps,prb,reprint,superscriptaddress,showpacs]{revtex4-1}
\usepackage{graphicx}
\usepackage{color}
\usepackage{amsbsy}
\usepackage{amsmath}
\usepackage[normalem]{ulem}
\usepackage{lipsum}
\usepackage[rightcaption]{sidecap}
\usepackage{setspace}

\begin{document}

\title{\fontfamily{phv}\selectfont{\bfseries{Raman and fluorescence characteristics of resonant inelastic X-ray scattering from doped superconducting cuprates}}}
 

\author{H. Y. Huang}
\affiliation{National Synchrotron Radiation Research Center, Hsinchu
30076, Taiwan} \affiliation{Program of Science and Technology of Synchrotron Light Source, National Tsing Hua University, Hsinchu 30013, Taiwan}

\author{C. J. Jia}
\affiliation{SIMES, SLAC National Accelerator Laboratory, Menlo Park, California 94025, USA}

\author{Z. Y. Chen}
\affiliation{Department of Physics, National Tsing
Hua University, Hsinchu 30013, Taiwan}

\author{K. Wohlfeld}
\affiliation{Institute of Theoretical Physics, University of Warsaw, PL-02093 Warsaw, Poland}

\author{B. Moritz}
\author{ T. P. Devereaux} 
\affiliation{SIMES, SLAC National Accelerator Laboratory, Menlo Park, California 94025, USA}

\author{W. B. Wu}\author{J. Okamoto}
\affiliation{National Synchrotron Radiation Research Center, Hsinchu 30076, Taiwan}

\author{W. S. Lee}  \author{M. Hashimoto} \affiliation{SIMES, SLAC National Accelerator Laboratory, Menlo Park, California 94025, USA}

\author{Y. He} \affiliation{SIMES, SLAC National Accelerator Laboratory, Menlo Park, California 94025, USA}
\affiliation{Department of Applied Physics, Stanford University, Stanford, California 94305, USA}

 \author{Z. X. Shen} \affiliation{SIMES, SLAC National Accelerator Laboratory, Menlo Park, California 94025, USA}
\affiliation{Department of Applied Physics, Stanford University, Stanford, California 94305, USA}
\affiliation{Department of Physics, Stanford University, Stanford, California 94305, USA}

\author{Y. Yoshida} \author{H. Eisaki} 
\affiliation{Nanoelectronics Research Institute, National Institute of Advanced Industrial Science and Technology, Tsukuba, Ibaraki 305-8562, Japan}

\author{C. Y. Mou} \affiliation{Department of Physics, National Tsing Hua University, Hsinchu 30013, Taiwan}

\author{C. T. Chen}
\affiliation{National Synchrotron Radiation Research Center, Hsinchu 30076, Taiwan}

\author{D. J. Huang}
\affiliation{National Synchrotron Radiation Research Center, Hsinchu 30076, Taiwan} \affiliation{Department of Physics, National Tsing Hua University, Hsinchu 30013, Taiwan}

\date{\today}

\begin{abstract}
Measurements of spin excitations are essential for an understanding of spin-mediated pairing for superconductivity; and resonant inelastic X-ray scattering (RIXS) provides a considerable opportunity to probe high-energy spin excitations. However, whether RIXS correctly measures the collective spin excitations of doped superconducting cuprates remains under debate. Here we demonstrate distinct Raman- and fluorescence-like RIXS excitations of Bi$_{1.5}$Pb$_{0.6}$Sr$_{1.54}$CaCu$_{2}$O$_{8+{\delta}}$ in the mid-infrared energy region. Combining photon-energy and momentum dependent RIXS measurements with theoretical calculations using exact diagonalization provides conclusive evidence that the Raman-like RIXS excitations correspond to collective spin excitations, which are magnons in the undoped Mott insulators and evolve into paramagnons in doped superconducting compounds. In contrast, the fluorescence-like shifts are due primarily to the continuum of particle-hole excitations in the charge channel. Our results show that under the proper experimental conditions RIXS indeed can be used to probe paramagnons in doped high-$T_c$ cuprate superconductors.
\end{abstract}

\maketitle


Inelastic X-ray scattering is a powerful spectroscopic tool to probe charge fluctuations and lattice dynamics of materials \cite{Kotani01,Ament11}. Because of core-level spin-orbit interactions in the intermediate
state, resonant inelastic X-ray scattering (RIXS) also can be sensitive to spin excitations of transition-metal compounds under the proper conditions \cite{deGroot98}. $L$-edge ($2p\rightarrow3d$)  RIXS  is increasingly becoming an important spectroscopic technique to examine the electronic dynamics of diverse materials. The Cu $L_3$-edge RIXS, which does not require large sample volumes and can access relatively high energies, has been used to probe spin excitations in antiferromagnetic insulating \cite{Braicovich09,Schlappa09, Guarise10,Braicovich10a, Schlappa12}, hole-doped \cite{Guarise14,Minola15,Braicovich10b,Tacon13, Tacon11,Dean13a, Dean13b}, and electron-doped \cite{Lee14} cuprates.

The interpretation of these RIXS measurements in terms of a collective magnetic mode that evolves with doping from a coherent magnon in the undoped phase to a damped magnetic excitation--paramagnon--in the doped phase, however, is in dispute \cite{Jia14,Benjamin14,Guarise14,Minola15,Braicovich10b,Tacon13, Tacon11,Dean13a, Dean13b}. First, RIXS measurements have revealed the persistence of these excitations well beyond the antiferromagnetic phase of cuprates, even in heavily-doped compounds for which superconductivity disappears \cite{Braicovich10b,Tacon13, Tacon11,Dean13a, Dean13b}.
Calculations which employ small clusters demonstrate that the RIXS cross section in a cross polarized geometry can be approximated by the spin dynamical structure factor $S(\mathbf{q}, \omega)$ \cite{Jia14}. 
However, other techniques using a single-band model of noninteracting quasiparticles \cite{Benjamin14} seem to explain RIXS measurements on Bi$_{2}$Sr$_{2}$CaCu$_{2}$O$_{8+{\delta}}$ (Bi2212) \cite{Guarise14} without invoking collective magnetic excitations. Intriguingly, very recent RIXS measurements on YBa$_{2}$Cu$_{3}$O$_{6+x}$ provide evidence that the mid-infrared RIXS in doped cuprates predominantly arises from magnetic collective modes, but no results due to the continuum of particle-hole excitations are reported \cite{Minola15}.

To reconcile these seemingly contradictory findings and interpretations in doped superconducting cuprates,
we performed comprehensive RIXS measurements on Bi$_{1.5}$Pb$_{0.6}$Sr$_{1.54}$CaCu$_{2}$O$_{8+{\delta}}$ (Pb-Bi2212) together with theoretical analysis using small-cluster exact diagonalization (ED). As its electronic structure and spin excitations have been well studied, Bi2212 is an ideal system to test whether RIXS correctly measures collective magnetic excitations in doped cuprates. Here we show that under the proper experimental conditions RIXS indeed can be used to probe paramagnons in doped high-$T_c$ cuprate superconductors. Our results highlight the powerful role that X-ray polarization, incident photon energy, and mometum transfer associated with the RIXS process can have on identifying, characterizing, and disentangling the RIXS excitations in the mid-infrared energy region.

\vspace{3mm}
\noindent {\fontfamily{phv}\bfseries Results}

\begin{figure}[ht]
\includegraphics[width=0.9\columnwidth]{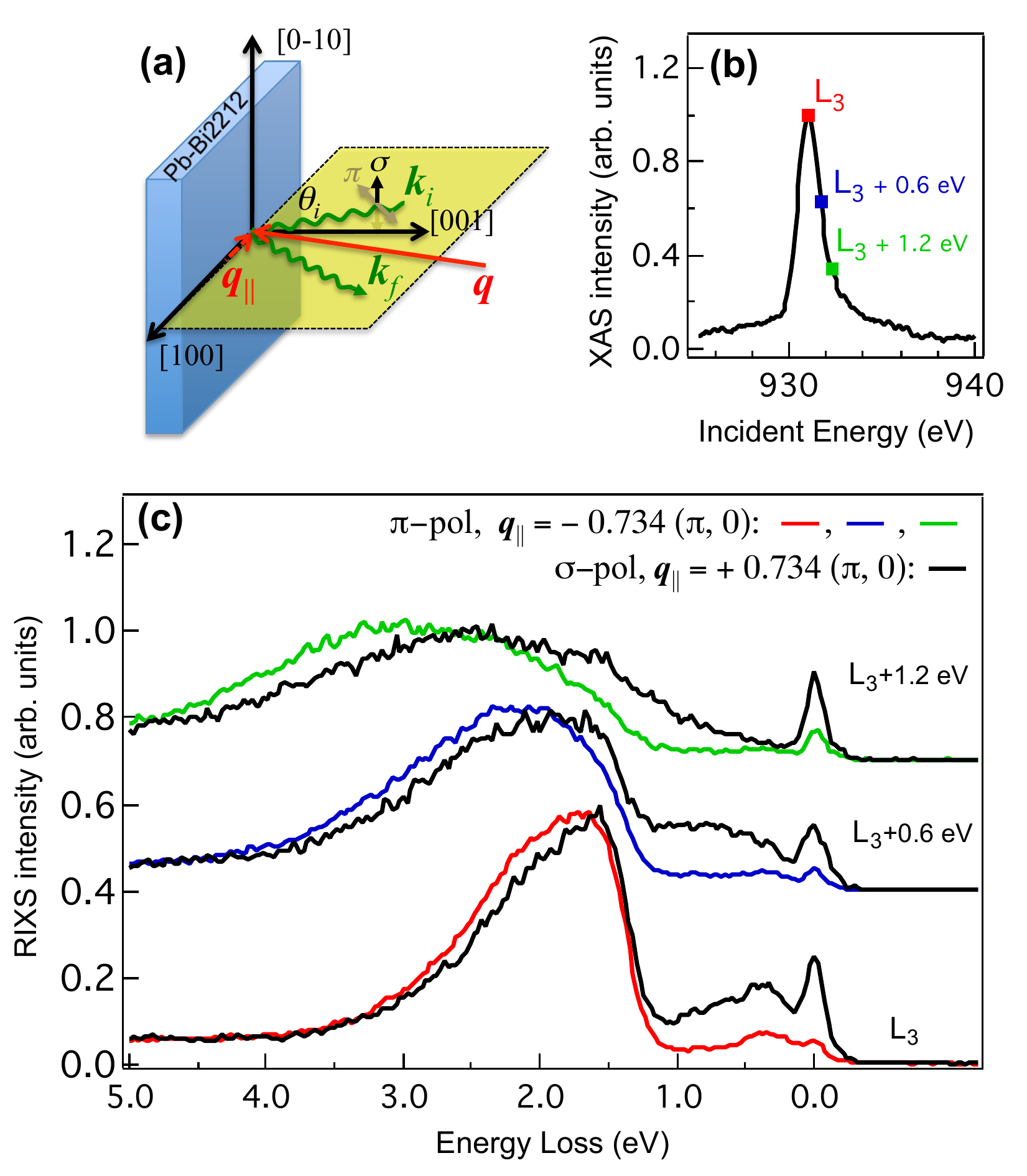}
\caption{{\bf RIXS spectra of Pb-Bi2212.}
({\bf a}) RIXS scattering geometry. The angle between the incident X-ray and the $ab$ plane of the sample is $\theta_{i}$.   The wave vectors of incident and scattered X-rays are ${\bf k}_i$ and ${\bf k}_f$, respectively. The momentum transfer is $\mathbf{q}= {\bf k}_{i} - {\bf k}_{f}$, and its projection onto the $ab$ plane is  $\mathbf{q}_\|$
({\bf b}) Cu $L_3$-edge X-ray absorption (XAS) measured in the fluorescent yield mode. The energy bandwidth of incident X-ray is 0.6 eV. ({\bf c}) RIXS spectra with incident X-ray energy set to the $L_3$ absorption threshold, 0.6~eV above, and 1.2~eV above using $\pi$-polarized (color) and $\sigma$-polarized (black) X-rays with $\theta_{i}=110^{\circ}$ and and $20^{\circ}$, i.e. grazing-exit $\mathbf{q}_{\|} = -0.734(\pi, 0)$ and grazing-incidence $\mathbf{q}_{\|} =  0.734(\pi, 0)$, respectively.  Spectra are normalized to the $dd$ excitations and also offset for clarity. } \label{Fig1}
\end{figure}

\noindent {\bf RIXS spectra of Pb-Bi2212}. Figure~\ref{Fig1} plots the incident photon-polarization dependence of RIXS spectra taken at 20~K, well below the superconducting transition temperature T$_{c} = 95$~K. 
The polarization of the incident X-ray is within or perpendicular to  the scattering plane, i.e. $\pi$ or $\sigma$ polarization, respectively.  
Figure~\ref{Fig1}(c) shows RIXS spectra excited with $\pi$- and $\sigma$-polarized  X-rays of energy set to the $L_3$ absorption threshold, 0.6~eV above, and 1.2~eV above, as indicated in Fig.~\ref{Fig1}(b).  
The RIXS intensities for energy loss between 1 and 3~eV arise mostly from $dd$ excitations in which the Cu $3d_{x^{2}-y^{2}}$ hole has been excited into previously filled $3d$ orbitals. 
Between 0 and 1~eV, the RIXS spectra are sensitive to incident photon energy and polarization, and in fact RIXS spectra of $\pi$ and $\sigma$ polarization show drastically different dependences on incident photon energy, particularly in the mid-infrared region.

\begin{figure}[ht]
\includegraphics[width=0.9\columnwidth]{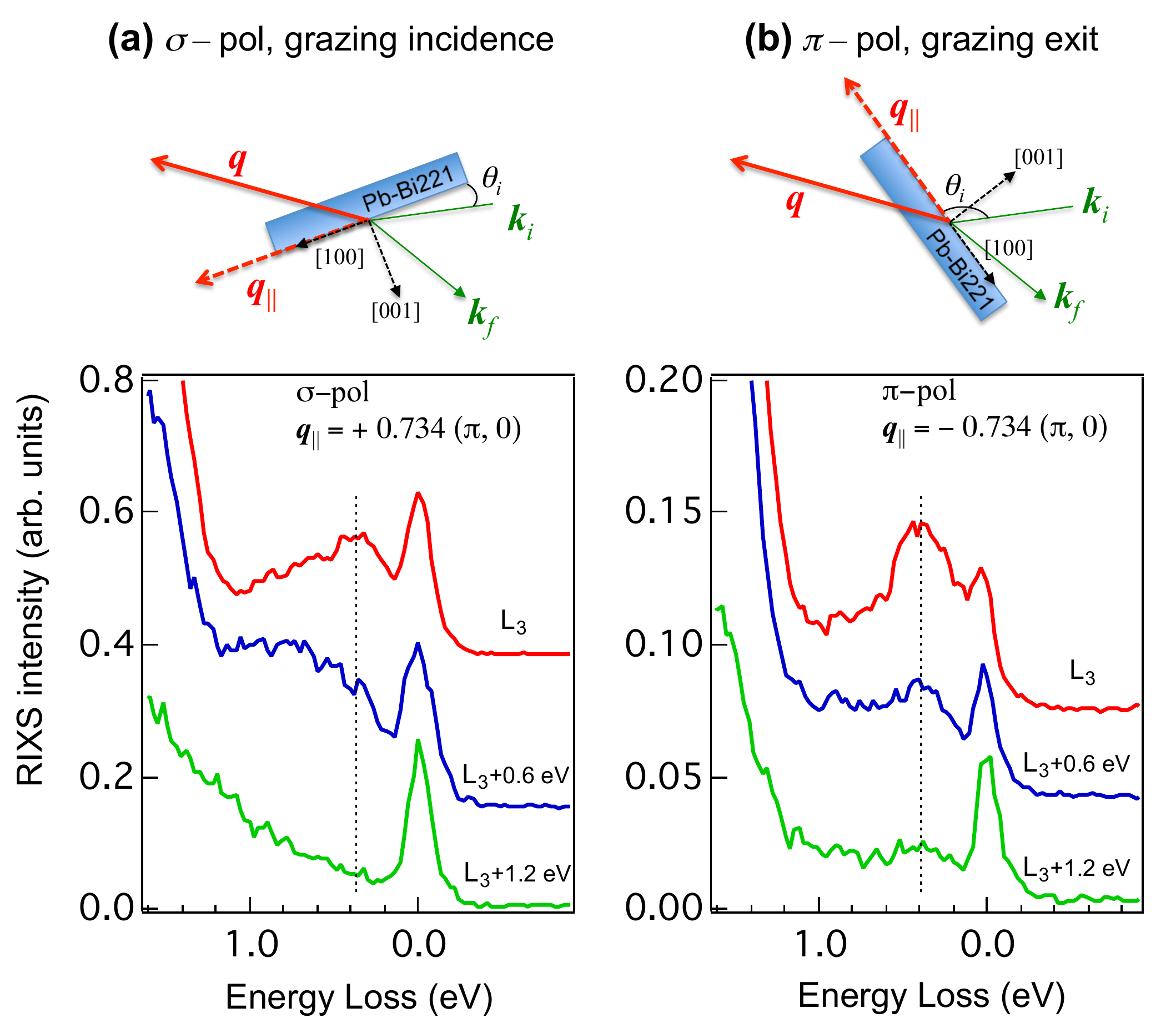}
\caption{{\bf Magnified plots of RIXS excited with X-rays at selected energies.} 
({\bf a}) \& ({\bf b}): RIXS spectra taken with $\sigma$-polarized incident X-rays under a grazing-incidence geometry and $\pi$-polarized incident X-rays under a grazing-exit geometry, respectively. The RIXS spectra of each scattering geometry are normalized to the elastic scattering of energy set to the $L_3$ absorption threshold. The RIXS spectra are vertically offset for clarity. Top panels illustrate the scattering geometries in which the scattering plane is defined by the [001] \& [100] of the Pb-Bi2212 crystal.    All notations are defined in Fig. \ref{Fig1}. The dashed lines serve as guides to the eyes.} \label{Fig2}
\end{figure}

\noindent {\bf Distinct Raman- and fluorescence-like RIXS.}
The detection of magnetic RIXS is known to be dependent on the polarization of incident photons. According to single Cu$^{2+}$ ion calculations \cite{Dean13b,Jia14,Braicovich10a}, the  cross section for a single spin flip is enhanced for grazing-incidence geometry ($q_{\|} > 0$) with incident X-rays of $\sigma$-polarization and grazing-exit geometry ($q_{\|} < 0$) with  incident X-rays of $\pi$-polarization. 
Figure \ref{Fig2} shows magnified plots of RIXS recorded under theses conditions.
With grazing-incidence X-rays of $\sigma$-polarization, we found that the energy loss centroid of the broad mid-infrared RIXS feature, as plotted in Fig. \ref{Fig2}(a), shifts with increasing incident energy from the $L_3$ edge to 0.6~eV and 1.2~eV above it.  Conversely, as shown in Fig. \ref{Fig2}(b), the RIXS intensity at an  energy loss of about 300-350~meV using $\pi$-polarized X-rays does not shift with the incident photon energy, indicating its Raman-like nature.

The photon energy dependence of the $\sigma$ spectra is consistent with the previous RIXS measurement on the same compound \cite{Guarise14}. This fluorescence-like feature was explained as particle-hole excitations with a single band model \cite{Benjamin14}. 
Such particle-hole interpretations may not hold in magnetic RIXS, perhaps arising from the charge channel. 
In fact, at the Cu $L$-edge, the maximum energies of the RIXS excitations with both $\sigma$ ($q_{\|} = +0.734$) and $\pi$ ($q_{\|} = -0.734$) polarizations coincide. However, their incident-photon-energy dependences are drastically different, indicating that at least two types of excitations coexist in the energy range where ``paramagnons" in question reside.

\begin{figure}[h]
\includegraphics [width=0.95\columnwidth] {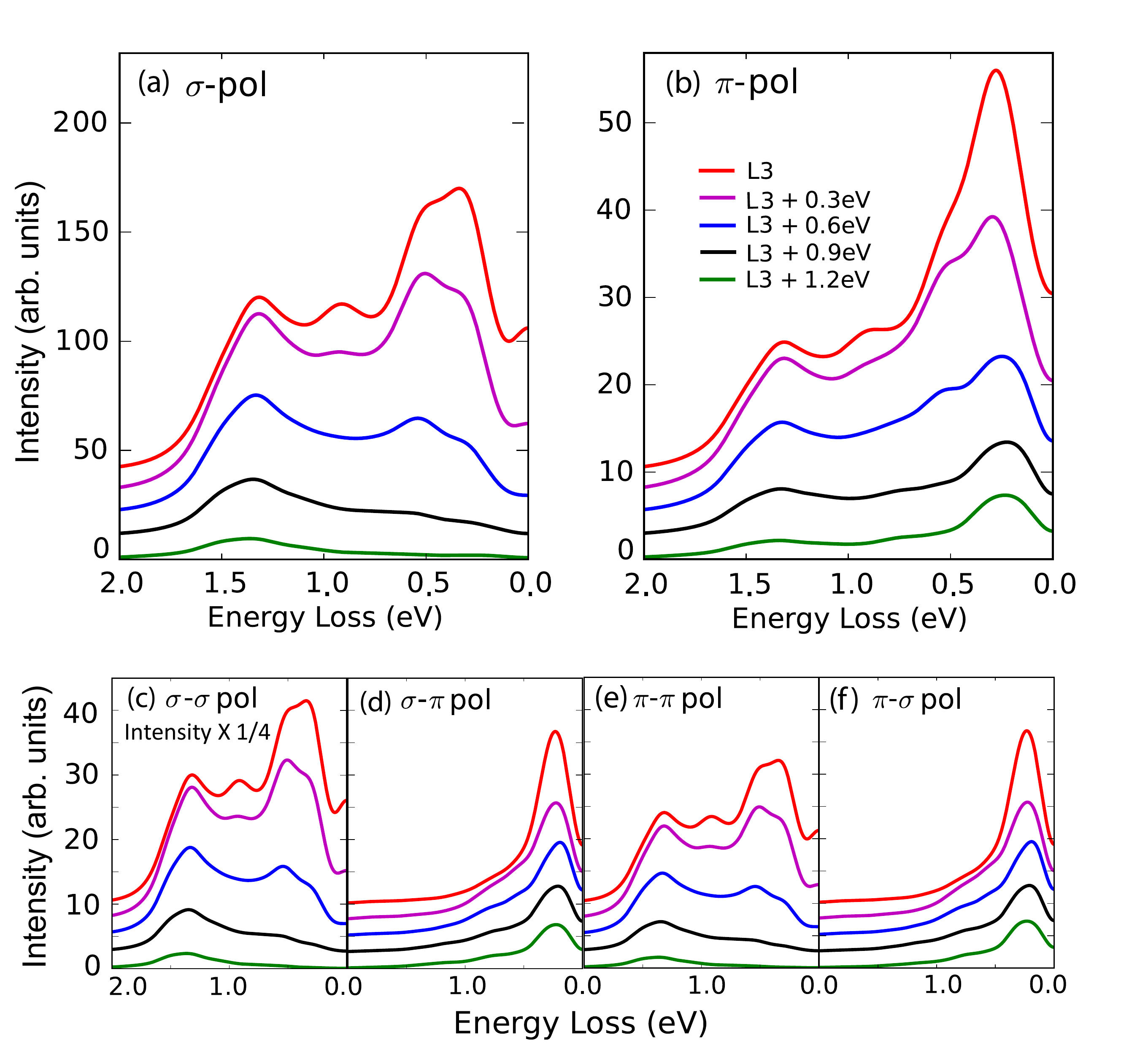}
\caption{{\bf Full RIXS ED calculations.} The ED calculations were performed as described in Methods using the single-band Hubbard model on a 12-site cluster under TABC (100 boundary conditions were used) at 16.7\% hole-doping. The parameter values are given in Methods. Panel (a) and (b) show the calculated spectra for $\sigma$-incoming polarization and $\pi$-incoming polarization respectively. Panel (c) through (f) show the calculated spectra also discriminating the outgoing photon polarization, which are plotted on the same linear intensity scale (intensity of Fig 3(c) is multiplied by 1/4). For all panels an offset is used for clarity. The transferred momentum is (2$\pi$/3, 0) for $\sigma$-polarization incidence and (-2$\pi$/3, 0) for $\pi$-polarization incidence in theory units where the lattice constant $a$ is 1. }\label{ED}
\end{figure}

\vspace{3mm}
\noindent {\bf ED Calculations of RIXS spectra.} We compare our measurements with ED calculations for the RIXS spectra of the single-band Hubbard model (see Methods). This incorporates effects from full many-body interactions, core-level spin-orbit coupling, and the polarization of incident and scattered X-rays \cite{Jia14}, which allows us to cleanly distinguish between charge channel and magnetic excitations. The cross polarization channels $\pi-\sigma$ and $\sigma-\pi$ highlight single spin-flip magnetic excitations, while the $\pi-\pi$ and $\sigma-\sigma$ parallel polarization channels should be dominated by charge excitations or bimagnon-like spin excitations. 

Figures~\ref{ED}(a)~and~\ref{ED}(b) show the calculated spectra for $\sigma$-incoming and $\pi$-incoming polarization respectively, under similar scattering geometry to the experimental data shown in Fig. \ref{Fig1}, but with $\mathbf{q}_{\|}=(2\pi/3, 0)$. With parallel polarization, i.e. Figs. \ref{ED}(c) and \ref{ED}(e), the observed charge channel excitations at the absorption edge are broad, extending over a wide energy range from less than 0.5 eV up to above 1.5 eV. As a function of incident photon energy, the intensity of these excitations decreases, while the centroid moves to higher energies. These fluorescence-like excitations are associated with broad, less well-defined particle-hole excitations in the charge continuum, which in fact may be well captured by non-interacting or the random-phase approximation (RPA) treatments. 

In contrast, the excitations observed with cross polarization, i.e. Figs.~\ref{ED}(d) and \ref{ED}(f), are sharp, more well-defined single spin-flip excitations at an energy~$\sim$~300~meV. Increasing the incident photon energy above the absorption edge reduces the intensity of these excitations, while the energy loss remains fixed. These Raman-like excitations previously have been identified as single-magnons in calculations for undoped systems or paramagnons in calculations of the doped Hubbard model based on the character of the final states and their momentum dependence.

These results demonstrate that neither $\sigma$-polarized nor $\pi$-polarized incident X-rays perfectly discriminate between Raman-like spin or fluorescence-like charge excitations. However, differences in matrix elements mean that $\pi$-polarization enhances the magnetic response relative to that in the charge channel, leading to observations dominated by the Raman-like spin excitations. On the other hand, while the grazing-incidence geometry enhances spin-flip excitations for $\sigma$-polarization, the charge channel fluorescence-like excitations remain dominant due to the much larger matrix elements (see Fig. \ref{ED}(c) and note the rescaling factor).

\begin{figure}[h]
\includegraphics[width=0.8\columnwidth]{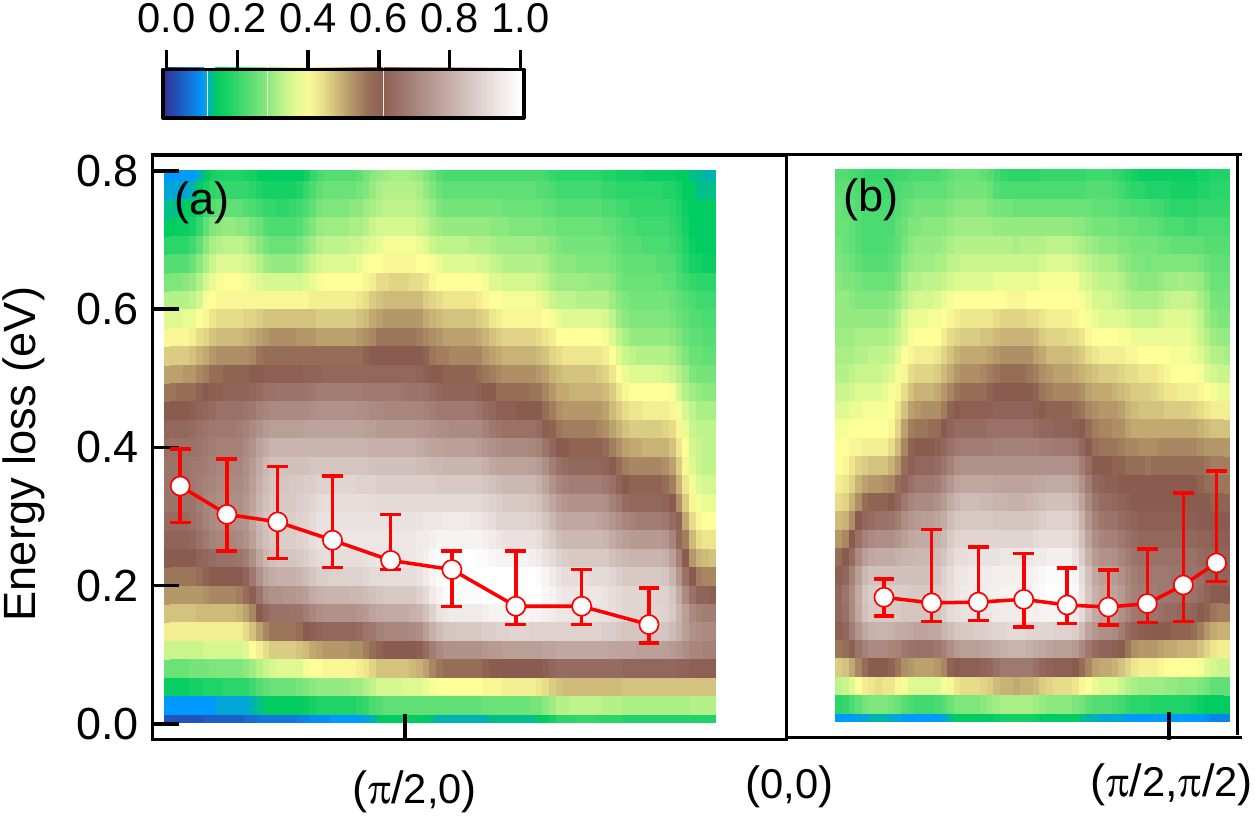}
\caption{{\bf Intensity maps of magnetic RIXS and dispersion of magnetic excitations in Pb-Bi2212.} The momentum transfer $\mathbf{q}_\|$ is varied along $(0, 0) \rightarrow (\pi, 0)$ in panel ({\bf a}) and $(0, 0) \rightarrow (\pi, \pi)$ in panel ({\bf b}). The RIXS intensities shown in the color maps are after background subtraction.  The excitation energies shown in red open circles are deduced from RIXS data with $\pi$-polarized incident X-rays as described in the text.} \label{Fig_disp}
\end{figure}

\vspace{3mm}
\noindent {\bf Dispersion of paramagnons.} RIXS measurements with  $\pi$-polarization for the broad spin excitations in Pb-Bi2212 confirm the character of collective excitations.  We performed momentum-dependent measurements with $\mathbf{q}_{\|}$ along the nodal and antinodal directions using $\pi$-polarized incident X-rays to highlight magnetic excitations. 
See Supplementary Figs.~S1~\&~S2 for the raw data and Methods for the data analysis.
Figure \ref{Fig_disp} shows intensity maps and dispersions of high-energy magnetic excitations after background subtraction. 
The RIXS profile of a paramagnon excitation can not be fitted fully with a Lorentzian or an antisymmetrized Lorentzian lineshape \cite{Dean13b,Minola15}.  For RIXS spectra with energy loss below 1~eV, we subtracted the elastic line and the background tail of the $dd$ excitations from the original spectra to obtain the magnetic RIXS spectra. The energy maximum after background subtraction is taken as the energy of the paramagnon excitation. We found that the paramagnon excitation along the nodal direction is less dispersive than that along the antinodal direction, and is softened near the antiferromagnetic zone boundary.

\vspace{3mm}
\noindent {\fontfamily{phv}\bfseries Discussion}

\noindent For incoming and scattered X-rays of given polarizations and energies, one can apply the Kramers-Heisenberg formula to obtain the RIXS cross section  (see Eq.~(\ref{operator}) in Methods).
It is generally agreed that, due to the large spin-orbit coupling for the core hole created in the intermediate state of the RIXS process, a $3d$ electron in the final state can undergo a spin-flip with respect to that of the initial state \cite{Benjamin14,deGroot98,Haverkort10,Ament09,Marra12}.  However, some details of this process remain unclear.

In the scenario of collective spin excitations, one assumes that the  energy loss and momentum transfer between the incident and scattered  X-rays will correspond (under particular experimental conditions) to the energy and wave vector of the spin excitation. In this scenario the energy loss as a function of momentum reveals a distinctive dispersive feature (a ``magnon" or ``paramagnon"), independent of incident photon energy. That is, both at and above the $L_3$ absorption threshold, the energy loss is constant for a fixed momentum transfer, and therefore would be Raman-like.
This is supported by the fact that RIXS is ``proportional" to the spin dynamical structure factor, which, for the doped Hubbard model, posses a distinctive dispersion, resembling the  well-known ``magnon" of the undoped Hubbard model from the $\Gamma$ point toward the antiferromagnetic zone boundary \cite{Jia14}.

In the noninteracting electron-hole scenario,  the energy of the intermediate state of RIXS increases with incident photon energy; and the final state falls into the continuum of electron-hole excitations. Hence the spin-flip excitation is no longer at constant energy, but rather the resonant energy shifts toward higher energies as the incident photon energy transcends the $L_3$ absorption threshold, like a fluorescence feature in a RIXS spectrum \cite{Benjamin14}. The ED calculations show that the observed fluorescence-like shifts for $\sigma$-polarization are due primarily to the continuum of particle-hole excitations in the charge channel. 

Here we show that, subject to the experimental conditions of RIXS, signatures of both scenarios can be partially observed in the RIXS spectra. We observed two distinct types of RIXS spectra, Raman-like and fluorescence-like when the incident X-rays are $\pi$- and $\sigma$-polarized, respectively.   Unlike particle-hole excitations, the collective spin excitations probed by  $\pi$-polarized RIXS are sharp and well defined, not decaying into a continuum, and hence the energy loss remains unchanged. 
The competition between these two types of RIXS excitations will give rise to an opportunity to probe Mott-like versus band-like electronic states, analogous to the competition between resonance photoemission and Auger emission \cite{Suzuki13}. 

Our combined results, including photon-energy and momentum dependent RIXS with full ED calculations, consolidate the evidence for Raman-like collective spin excitations in RIXS measurements on hole-doped superconducting cuprates.
For measurements with  $\pi$-polarization and grazing-exit geometry shown in Fig.~\ref{Fig_disp}, the dispersion appear to resemble those measured with $\sigma$-polarization and grazing-incidence geometry \cite{Guarise14}.  Importantly, our measurements confirm the ``collapse" of the paramagnons, i.e. the notably reduced bandwidth along the nodal direction. This is indeed the behavior of Raman-like paramagnon excitations, rather than particle-hole excitations with trivial spin flips; therefore, the ``collapse" of the paramagnon remains an open question that cannot be simply explained by non-interacting single-band theories.

\vspace{3mm}
\noindent {\fontfamily{phv}\bfseries Methods}

\noindent {\bf Experimental details.} We measured RIXS on a Bi$_{1.5}$Pb$_{0.6}$Sr$_{1.54}$CaCu$_{2}$O$_{8+{\delta}}$ (Pb-Bi2212) single crystal at selected incident photon energies and polarizations using the AGM-AGS spectrometer at beamline 05A1 of the National Synchrotron Radiation Research Center (NSRRC), Taiwan \cite{Lai14}. 
To measure the inelastic scattering arising from spin excitations, we set the photon energy about the $L_3$-edge absorption of Cu, i.e. 931~eV. RIXS spectra were recorded with a combined energy resolution $\sim$ 125~meV without analysis of the polarization of scattered X-rays.  
The polarization of the incident X-ray was within or perpendicular to  the scattering plane, i.e. $\pi$ and $\sigma$ polarizations, respectively. 

Being among the most studied cuprate superconductors, Bi2212 cuprate shares a two-dimensional layered perovskite structure with superconductivity taking place in the copper-oxide planes. 
The optimally doped Pb-Bi2212 single crystal of high quality was grown with the floating-zone method. The concentration of holes was optimized on annealing the samples under flowing N$_2$. Pb doping in Pb-Bi2212 suppresses the incommensurate structural modulation of the BiO planes along the pseudo-tetragonal axis \cite{Hashimoto10,He11}. The Pb-Bi2212 crystal was cleaved in air before being loaded into the vacuum chamber ($\sim 1\times10^{-8}$ mbar) for RIXS measurements, which were taken on the sample at 20~K, well below the superconducting transition temperature T$_{c} = 95$~K. 

As illustrated in Fig.~\ref{Fig1}(a), we varied the incident angle $\theta_i$ to change the momentum transfer of RIXS projected onto the $ab$ plane, i.e.~$\mathbf{q}_\|$, while the scattering angle between the incident and scattered X-rays was fixed at 130$^\circ$. The Pb-Bi2212 crystal was rotated about its surface normal [001] in vacuum to select $\mathbf{q}_\|$ along the nodal (diagonal) or antinodal (parallel to the Cu-O bonds) direction in momentum space. 
Supplementary Figure S1 depicts original momentum-dependent RIXS spectra within an energy loss below 4~eV. To identify the excitation energy below 1~eV, we subtracted the elastic line and the background tail of the $dd$ excitations from the original spectra, as demonstrated in Supplementary Fig. S2(a). The energy of the paramagnon excitation in each spectrum is determined by the maximum of the broad magnetic RIXS after background subtraction. The energy range in which the RIXS intensity is larger than 95\% of the maximum intensity is taken as the error bar of the paramagnon energy. Supplementary Figures S2(b) and S2(c) plot RIXS spectra after the aforementioned background subtraction within the energy loss below 0.9 eV. 

\vspace{3mm}
\noindent{\bf Full RIXS ED calculations.} The Cu $L_3$-edge RIXS cross-section is calculated using the Kramers-Heisenberg formula for the single-band Hubbard model 

\begin{widetext}

\begin{equation} 
\begin{aligned}
&I(\mathbf{q}, \omega, \omega_{in})
= \sum_{f} | \langle f | R_{\omega_{in}}^{\boldsymbol{\epsilon}_i \boldsymbol{\epsilon}_o} | 0 \rangle | ^ 2 \delta (\omega + E_0 - H_{Hubbard} )\\
& R_{\omega_{in}}^{\boldsymbol{\epsilon}_i \boldsymbol{\epsilon}_o}  = \sum_{j\alpha\sigma\sigma^{\prime}} e^{i\mathbf{q}\cdot\mathbf{R}_{j}} \mathit{T^{\boldsymbol{\epsilon}_o \dagger}_{j\alpha\sigma^{\prime}}} \frac{1}{\omega_{in} +\mathit{E}_0 - \mathit{H}_\mathrm{Hubbard} - \mathit{H}_{\rm CH} + \mathit{i}\Gamma} \mathit{T^{\boldsymbol{\epsilon}_i}_{j\alpha\sigma}} \\
\end{aligned}\label{operator}
\end{equation} 
in which
\begin{equation} 
\begin{aligned}
&\mathit{H_\mathrm{Hubbard}}= -t \sum_{<ij>,\sigma} d_{i\sigma}^{\dagger}d_{j\sigma}-t^{\prime} \sum_{\ll ij\gg,\sigma} d_{i\sigma}^{\dagger}d_{j\sigma}+\sum_{i} U n_{i\uparrow}^d n_{i\downarrow}^d \\
&\mathit{H}_{\rm CH}=\sum_{i\alpha\sigma}(\epsilon^d-\epsilon^p)(1-n_{i\alpha\sigma}^p)
-\mathit{U}_c\sum_{i\alpha\sigma\sigma^{\prime}} n_{i\sigma}^d (1-n_{i\alpha\sigma^{\prime}}^p) +\lambda\sum_{i\alpha\alpha^{\prime} \sigma\sigma^{\prime}} p_{i\alpha\sigma} ^{\dagger} \langle p_{\alpha\sigma}|\mathbf{l}\cdot \mathbf{s} |p_{\alpha^{\prime}\sigma^{\prime}}\rangle p_{i\alpha^{\prime}\sigma^{\prime}},
\end{aligned}
\end{equation} 
\end{widetext}
where $\mathbf{q}$ is the momentum transfer; $\omega_{in}$ and ${\omega = \omega_{in} - \omega_{out}}$ are the incident photon energy (in our study the Cu $L_3$-edge) and photon energy transfer, respectively; $\mathit{E}_0$ is the ground state energy of $H_{\rm Hubbard}$; $| 0 \rangle$ is the ground state wave 
function of $H_{\rm Hubbard}$; $| f \rangle$ is the final state wave function of $H_{\rm Hubbard}$; ${\mathit{T^{\boldsymbol{\epsilon}}_{i\alpha\sigma}} = \langle d_{x^2-y^2, \sigma} | \hat{\epsilon} \cdot \hat{r} | p_{\alpha\sigma} \rangle d^{\dagger}_{i\sigma}p_{i\alpha\sigma}}$ (and h.c.) dictates the dipole transition process 
from Cu $2p$ to the $3d$ level (or from Cu $3d$ to $2p$), with the X-ray polarization $\boldsymbol{\epsilon}$ ($\boldsymbol{\epsilon}_i$ for incoming photon and $\boldsymbol{\epsilon}_o$ for outgoing photon) either $\pi$ or $\sigma$; and 
$\Gamma$ is the inverse core-hole lifetime. In $\mathit{H_\mathrm{\rm Hubbard}}$, {$<...>$} and {$\ll ... \gg$} represent a sum over the nearest and next nearest neighbor sites, respectively. The Hamiltonian for the intermediate state also involves the on-site energy $\epsilon^d-\epsilon^p$ for creating a $2p$ 
core hole, Coulomb interaction $U_c$ induced by the core-hole and spin-orbit coupling $\lambda$, all denoted as in $\mathit{H}_{\rm CH}$. 

In the RIXS process, the optical transition operator $T^{\boldsymbol{\epsilon}_i}$ excites the ground state $|0\rangle$ to an intermediate sate $T^{\boldsymbol{\epsilon}_i}|0\rangle$ through creation of a core hole with lifetime broadening $\Gamma$ reflected in the spectra, while promoting an electron to the $3d$ shell. After a fluctuation described by the propagator $(\omega_{i}+E_{i} - \mathit{H}_\mathrm{Hubbard} - \mathit{H}_{\rm CH} + i{\Gamma})^{-1}$  the operator $T^{\boldsymbol{\epsilon}_{o} \dag}$ fills the core hole and at the same time annihilates a $3d$ electron.
The parameters used in the RIXS calculation are  $t=0.4$ eV, $U = 8t=3.2$ eV, $t^{\prime}=-0.3t=-0.12$ eV, $\epsilon^d-\epsilon^p=930$ eV, $U_c = -6t = -2.4$ eV, $\lambda = 13$ eV and $\Gamma=1t = 0.4$ eV. The RIXS results were obtained for a Lorentzian broadening with half width at half maximum (HWHM) = $0.01$~eV ($0.025t$) and a Gaussian broadening with $\sigma$ = 0.08~eV ($0.2t$) on the energy transfer. The numerical calculations were performed on the 12-site cluster, using Parallel Arnoldi Package (PARPACK) \cite{PARPACK}, bi-conjugate gradient stabilized method (BiCGSTAB) \cite{BiCGSTAB} and continued fraction expansion method \cite{Dagotto94,Jia12}. 

To reduce finite size effects, we also implemented twisted average boundary condition (TABC) \cite{Gros96}.  The boundary condition is expressed as $d_{i+R_a} = e^{i \phi_a} d_i$ and $d_{i+R_b} = e^{i \phi_b} d_i$. For 100 boundary conditions used in our calculations, we set $\phi_a = 0, 2\pi/10,  2\pi/10 \times 2, ... , 2\pi/10 \times 9$ and $\phi_b = 0, 2\pi/10,  2\pi/10 \times 2, ... , 2\pi/10 \times 9$. The TABC results are obtained by calculating the RIXS spectra for $H_{\rm Hubbard}$ on the 12-site cluster with the 100 boundary conditions and averaging the results \cite{Tohyama04,Gros96}.


\vspace{3mm}
\noindent {\fontfamily{phv}\bfseries Acknowledgements}

\noindent We thank Shiwei Zhang, Atsushi Fujimori and Sumio Ishihara for insightful discussions and the staff of NSRRC for their technical help. This work was supported in part by the Ministry of Science and Technology of Taiwan under Grant No. 103-2112-M-213-008-MY3. Work at SLAC was supported by the US Department of Energy, Office of Science, Basic Energy Sciences, Materials Sciences and Engineering Division, under Contract DE-AC02-76SF00515. K. W. acknowledges support from the Polish National Science Center under Project No. 2012/04/A/ST3/00331.

\vspace{3mm}
\noindent {\fontfamily{phv}\bfseries Author contributions}

\noindent H.Y.H., Z.Y.C., W.B.W. and J.O. performed RIXS measurements and analyzed the data. C.J.J. performed ED calculations. C.T.C. designed the RIXS beamline and spectrometer. M.H., Y.H., Z.X.S., Y.Y. and H.E. synthesized and prepared the single-crystals used for RIXS measurements. H.Y.H., C.J.J., K.W., B.M., W.S.L., T.P.D., C.Y.M. and D.J.H. discussed the results and wrote the paper. D.J.H. is responsible for project planning.

\vspace{3mm}
\noindent {\fontfamily{phv}\bfseries Competing Financial Interest}: The authors declare no competing financial interests.

\vspace{3mm}
\noindent {\fontfamily{phv}\bfseries Correspondence} and requests for materials should be addressed to D.J.H. (email: djhuang@nsrrc.org.tw)

\newpage 
\widetext
\begin{center}
\textbf{\large Supplementary Information }
\end{center}
\setcounter{equation}{0}
\setcounter{figure}{0}
\setcounter{table}{0}
\setcounter{page}{1}
\makeatletter
\renewcommand{\figurename}{Supplementary Figure}
\renewcommand{\thefigure}{S\arabic{figure}}
\renewcommand{\bibnumfmt}[1]{[S#1]}
\renewcommand{\citenumfont}[1]{S#1}
\begin{figure}[ht]
\includegraphics[width=0.7\columnwidth]{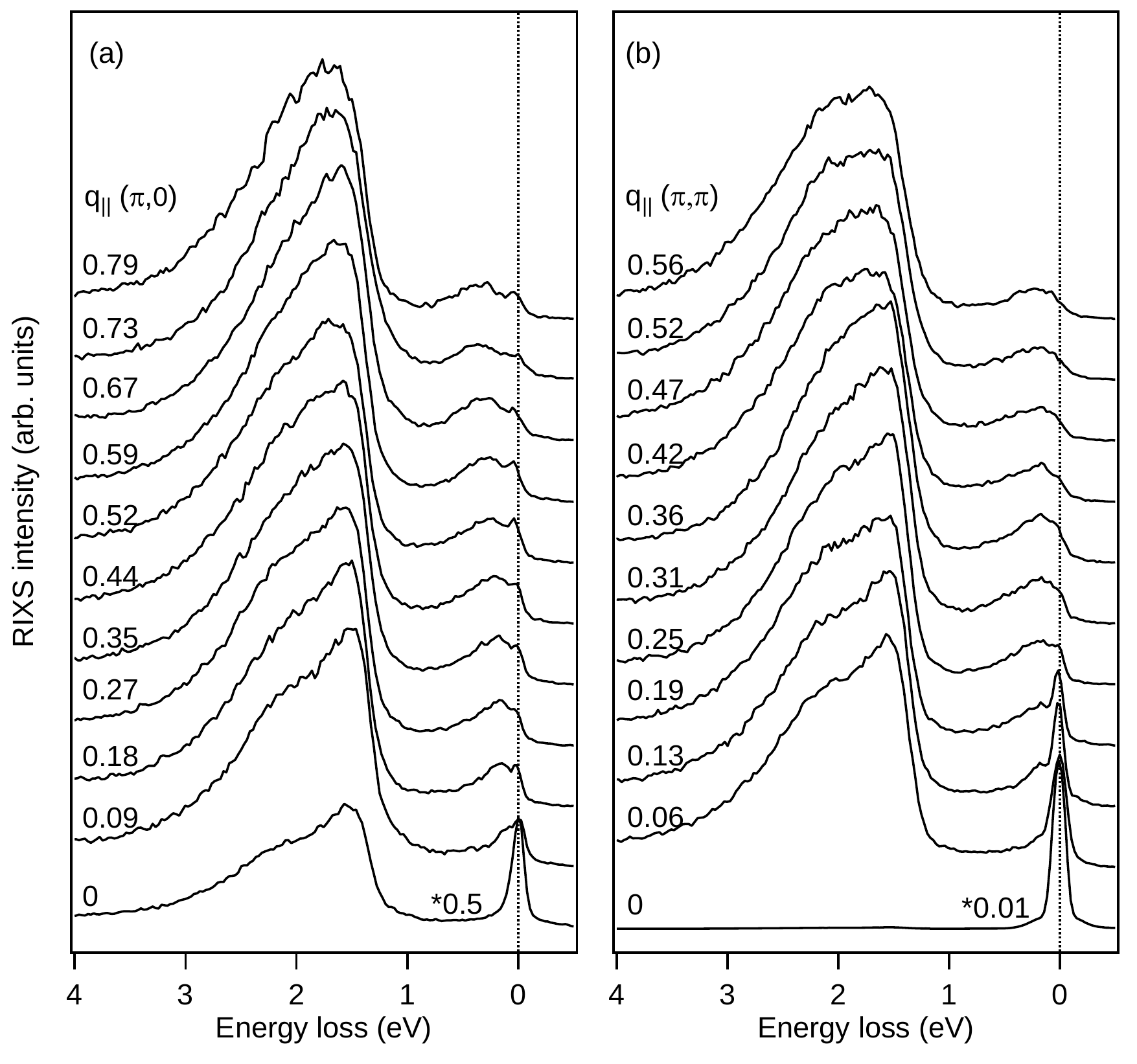}
\caption{{\bf Momentum-dependent $L_{3}$-edge RIXS spectra.} (a) \& (b) RIXS spectra along antinodal and nodal directions measured with $\pi$-polarized incident X-rays. $\mathbf{q}_\|$ is expressed as $q_\|$($\pi$,0) or $q_\|$($\pi$,$\pi$). The incident angle $\theta_{i}$ measured between the incident X-ray and sample surface is varied between $65^{\circ}$ at $q_\|=0$ to $115^{\circ}$ at the largest $q_\|$. All the data are normalized to the $dd$ excitations.} 
\end{figure}

\begin{figure}[ht]
\includegraphics[width=0.8\columnwidth]{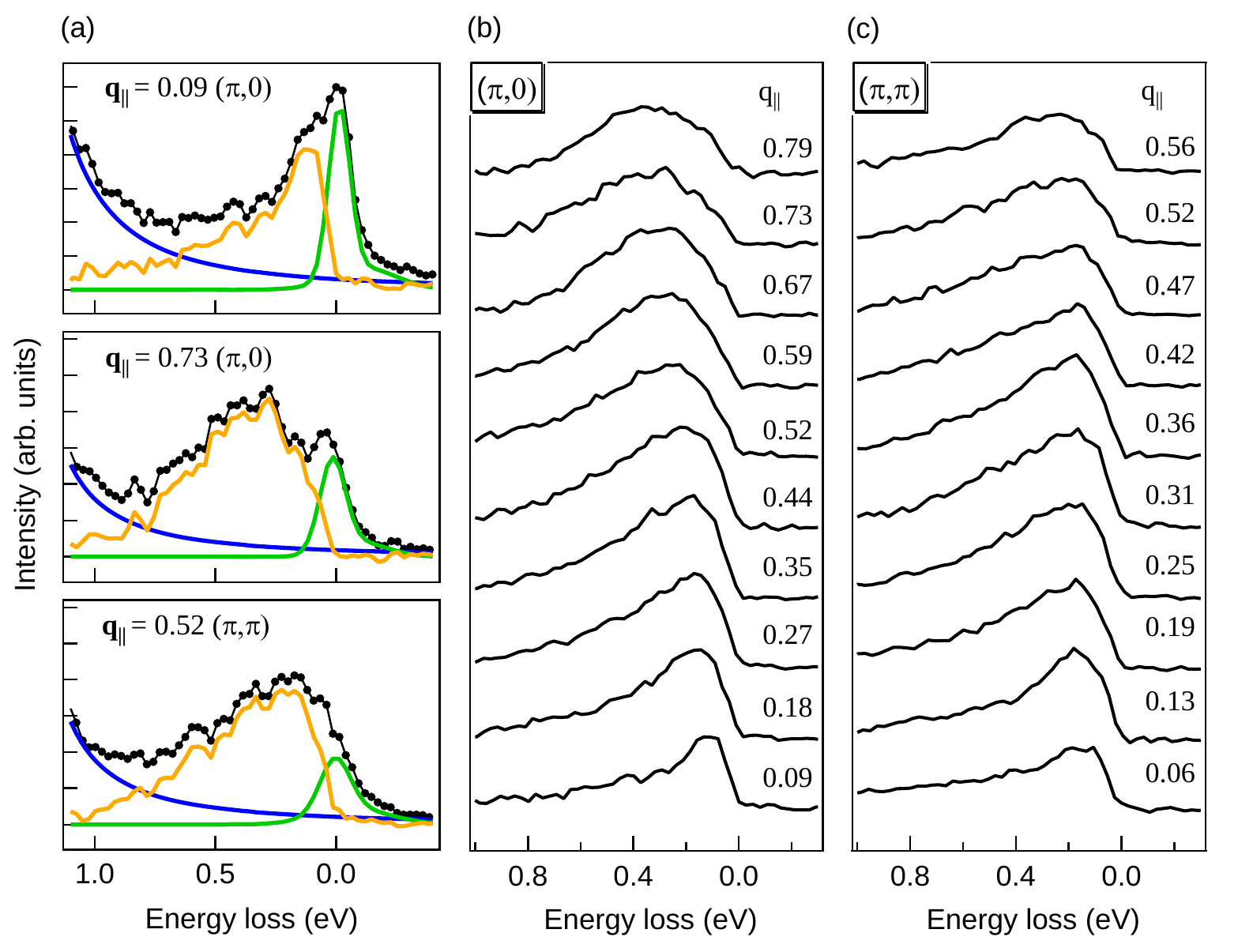}
\caption{{\bf Momentum-dependent $L_{3}$-edge RIXS spectra after subtraction of elastic peak.}
(a) Selected original RIXS data plotted together with the elastic line (green) and the background tail of the $dd$ excitation (blue) at selected $\mathbf{q}_\|$ along the antinodal ($\pi$,0) and nodal ($\pi$,$\pi$) directions. The data after background subtraction are plotted with orange lines. (b) \& (c) RIXS spectra of various $\mathbf{q}_\|$ after removing the elastic line and the background of $dd$ excitations.} 
\end{figure}

\end{document}